\documentclass[preprint]{aastex}
\usepackage{epsfig}

\begin{document}

\title{A Peculiar Flaring Episode of Cygnus X-1}
\author{Wei Cui, Yu-Xin Feng, and Mark Ertmer}
\affil{Department of Physics, Purdue University, West Lafayette, IN 47907}

\begin{abstract}

Recent monitoring of Cyg X-1 with {\em RXTE} revealed a period of intense 
flaring, which started in October of 2000 and lasted until March of 2001. 
The source exhibited some quite unusual behaviors during this period. The 
soft X-ray flux of the source went up and down three times on a timescale 
of about one month, as discovered by the ASM aboard RXTE, before finally 
returning to the normal level (of the hard state). The observed spectral 
and temporal X-ray properties of Cyg X-1 are mostly intermediate between 
the canonical hard and soft states. This is known previously for strong 
X-ray flares, however, we show that the source did enter a period that 
resembles, in many ways, a sustained soft state during the last of the 
three flares. We make detailed comparisons between this flare and the 
1996 state transition, in terms of the observed X-ray properties, such 
as flux--hardness correlation, X-ray spectrum, and power density spectrum. 
We point out the similarities and differences, and discuss possible 
implications of the results on our understanding of the phenomena of 
flares and state transitions associated with Cyg X-1.

\end{abstract}


\section{Introduction}

Cygnus X-1 has always been considered as an archetypical black hole 
candidate (BHC; reviews by Oda 1977, Liang \& Nolan 1984, and 
Tanaka \& Lewin 1995). 
Its observed spectral and temporal X--ray properties have, therefore, 
often been used to distinguish BHCs from their neutron star counterparts. 
Though flawed, this approach has resulted in the discovery of many BHCs 
whose candidacy is subsequently confirmed by dynamical measurements of 
the mass of the compact object.

Cyg X-1 shows two distinct spectral states: the hard state and the soft 
state. The source spends most of the time in the hard state where the 
soft X-ray (2-10 keV) luminosity is relatively low and the power-law 
X-ray spectrum relatively flat. Once every few years, Cyg X-1 undergoes 
a transition from the hard state to the soft state. It remains in the 
soft state for weeks to months before returning to the hard state. 
During such a spectral state transition, the power-law spectrum 
``pivots'' around 10--20 keV, causing a significant increase in the soft
X-ray flux but a decrease in the hard X-ray ($\gtrsim 20$ keV) (e.g.,
Zhang et al. 1997). The source has only occasionally been observed in 
the soft state (Oda 1977; Liang \& Nolan 1984; Cui et al. 1997a) and is 
thus not as well studied as in the hard state. Besides the difference 
in the spectral properties of the source between the two states, the 
temporal properties are also quite different. The power density spectrum 
(PDS) of Cyg X-1 in the hard state can be characterized by a white noise
component at low frequencies and a power-law component at high 
frequencies, with the characteristic ``break frequency'' in the range 
of $\sim$0.04-0.4 Hz 
(reviews by van der Klis 1995 and Cui 1998). In the soft state, however, 
the PDS is dominated by ``$1/f$'' noise (Cui et al. 1997a). 
For both states, the PDS shows additional features at a few Hz (e.g., 
Belloni \& Hasinger 1990; Cui et al. 1997a). Moreover, the X-ray 
flux of Cyg X-1 is known to vary on all timescales down to at least 
milliseconds; the X-ray variability seems stronger in the hard state 
than in the soft state. 

Cyg X-1 is also known to experience frequent X-ray flares. While most 
flares last for less than a day, some can last for weeks to months, 
during which additional short-duration flares may also occur. Studies 
have shown that at least some of the major flares are similar to 
state transitions (Pottschmidt et al. 2001; Feng, Ertmer, \& Cui, in 
preparation). Therefore, the distinction between major flares and state 
transition can sometimes be quite ambiguous. The X-ray flares of 
Cyg X-1 are poorly understood due to the lack of high-quality data. 
In this {\it Letter}, we present results from recent observations of 
Cyg~X-1 during a peculiar period of intense flaring activities with 
the {\it Rossi X-ray Timing Explorer} (RXTE). 

\section{All-Sky Monitor Observations}

Starting in October of 2000, a series of long, intense X-ray flares 
were observed of Cyg X-1 by the All-Sky Monitor (ASM) aboard RXTE. 
Fig.~1 (top panel) shows a portion of the ASM light curve of the source 
for this time period. Three major flares are clearly seen in the light 
curve. They all reached roughly the same peak flux and lasted for 
roughly the same length of time (about a month), although 
their time profiles are quite different. Fig.~1 (middle panel) also 
shows a time series of the ratio of the count rate in the 3--5 keV 
band to that in the 1.5--3 keV band. Such a hardness ratio provides a 
rough measure of the X-ray spectral shape of the source. The results
show that the spectrum of Cyg X-1 softens significantly during each of
the flares. The minimum value of the hardness ratio is very similar for
the first and last flares but seems higher for the middle flare (although
the coverage of the later part of the latter is relatively sparse). It 
is interesting to note that both the peak flux and the minimum hardness 
ratio reached by the flares are nearly identical to those of the 1996
state transition of Cyg X-1 (Cui et al. 1997a). The question is now 
whether there are any clear distinctions between a flare that we see 
here and a canonical state transition of Cyg X-1.

To shed more light on the issue, we examined the correlation between the
soft X-ray flux of the source and its spectral hardness, following the
work of Wen, Cui, \& Bradt (2001). Those authors demonstrated, based 
mostly on the ASM data, that the flux--hardness correlation is a reliable 
indicator of the soft state, certainly more reliable than the soft X-ray 
flux alone. They found that the flux and hardness ratio are strongly 
{\em correlated} for the soft state but are weakly {\em anti-correlated} 
for the hard state (also see Li, Feng, \& Chen 1999). The transition 
between the two states is a gradual and continuous process, as 
quantified by the correlation coefficient. We adopted the Spearman
rank-order method (Press et al. 1992) to investigate correlation between
the soft flux and spectral hardness of the source, again following
Wen et al. (2001). Briefly, both the rate and hardness-ratio time series 
(with 90 s time bins, as shown in Fig.~1) were first parameterized with 
their rank numbers. The light curves were then broken into intervals that 
contain the same number of data points (18 in our case). Note that the 
intervals are not necessarily of the same duration due to the presence of 
coverage gaps. For each interval, a correlation coefficient between the 
two parameterized light curves was computed. The bottom panel of Fig.~1 
shows the evolution of the coefficient throughout the flaring period. 

The results show a weak, negative correlation during non-flare intervals,
indicating that the source reached the hard state (Wen et al. 2001). 
The most striking feature is, however, an interval of positive correlation 
(except for a dip in the middle) during the last 
flare. Interestingly, the average value of the coefficient for this flare
is about the same as that for the 1996 state transition (see Wen et al. 
2001). Therefore, the last flare seems quite similar to the 1996 soft 
state. However, the other two flares behave quite differently. Although 
roughly the same positive coefficient is achieved at times during the 
first flare, the fluctuation is quite large. The middle flare seems to be 
intermediate between the hard and soft states. We speculate that flares 
and state transitions 
may differ only in a quantitative sense; the underlying physical 
process(es) may be common for these seemingly different phenomena. To
make progress on these issues, we obtained detailed X-ray properties of 
the source during the last flare, using data from pointed observations,
and compared the properties to those during the 1996 state transition.

\section{Pointed RXTE Observations}

During the flaring period, Cyg X-1 was frequently observed by the large
area detectors aboard RXTE, namely, the Proportional Counter Array (PCA),
which covers a nominal energy range of 2--60 keV, and the High Energy
X-ray Timing Experiment (HEXTE), which covers a nominal energy range of
15--250 keV. The times of the pointed observation are indicated by tick
marks in Fig.~1. These observations provided data for us to examine the
detailed spectral and timing properties of the source during the last 
flare and compare them to those observed of the 1996 state transition. 

\subsection{Spectral Analysis and Results}

For each observation, we constructed an X-ray spectrum of Cyg X-1 from 
the PCA and HEXTE data with {\em FTOOLS (v5.0)}, as well as the calibration 
files and background models that accompanied this release of the software.
To minimize calibration uncertainties associated with the PCA, we only
used data from the first xenon layer of each detection unit (which is
most accurately calibrated). The trade-off is that we lose spectral
coverage at energies above roughly 30 keV. On the other hand, we could
rely on the HEXTE data to extend the coverage to higher energies. At low
energies ($\lesssim 5$ keV), the calibration of the PCA is also relatively 
uncertain. To be conservative, therefore, we ignored data below about 
4.5 keV for spectral analyses.

For technical reasons, not all five detector units of the PCA were 
turned on for the observations. The number of units in use varied 
from one observation to the other; so did the specific set of detector 
units. For a given observation, we used {\em Standard 2}
data to produce a spectrum for each detector unit. We then derived a 
background spectrum for that unit and subtracted it from the 
observed spectrum to obtain a source spectrum for spectral modeling. 
Similarly, we constructed an overall spectrum, along with a background 
spectrum, for each of the two HEXTE clusters. The individual 
background-subtracted spectra were then modeled jointly in
{\em XSPEC (v11.0)}. We allowed the relative normalizations of the 
detector units to vary in the fits, to account for any slight difference 
among the units of each instrument and the known difference between the 
two instruments. We also added 1\% systematic uncertainty to the data.

To facilitate a quantitative comparison between the last flare observed
here and the 1996 state transition, we simply adopted the empirical 
model used by Cui et al. (1997a) to fit the spectra. However, we did 
not include a disk blackbody component, because, by ignoring data below 
$\sim$4.5 keV, we were no longer able to reliably constrain it. The 
model now consists of a broken power law with a high-energy rollover.
In order to achieve acceptable fits, however, we had to also include a 
Gaussian function in the model to account for residuals between 
5--8 keV for all observations. The centroid energy of the Gaussian 
component varies in the range of 6.3--6.6 keV, which seems to indicate 
the presence of a Fe K emission line. For the fits we also fixed the 
hydrogen column density at 
$6.2 \times 10^{21}\mbox{ }cm^{-2}$ (Schulz et al. 2001; 
Ba{\l}uci\'{n}ska \& Hasinger 1991), again due to the 
lack of sensitivity of the RXTE data at low energies. The best-fit 
parameters are listed in Table~1 for a selected sample of observations 
that cover the last flare (see Fig. 1 for the times of these 
observations). The uncertainties shown represent 90\% confidence 
intervals. Fig.~2 illustrates the quality of a representative fit.

The results show clear spectral evolution of the source over the last 
flare. The spectrum softens progressively as the source flux increases 
toward the peak of the flare and hardens as the flux decreases from 
the peak; the two transitional periods seem to mirror each other 
(although the coverage of the ``low-to-high'' transition is poorer). 
At the peak, the 
parameters of the third and fourth observations are nearly identical to 
those of the 1996 soft state (see the results of the fourth observation 
in Table 2 of Cui et al. 1997a). In fact, the spectrum of the source is
similarly soft during all observations between (and around) these two
selected ones (see Fig. 1 for the times of other observations). 
Spectrally, this flare is, therefore, very similar to the 1996 state 
transition. However, one difference is noticeable. During the 1996 
episode, the flux-hardness correlation stayed positive throughout the
soft state (although the coverage was not very good; Cui et al. 1997b;
Wen et al. 2001). During the flare here, even when the flux-hardness
correlation is positive (see Fig.~1), the spectrum of the source is at 
times significantly harder (e.g., comparing the second and fifth 
observations with the third and fourth observations in Table 1).

\subsection{Timing analysis and Results}

We also proceeded to examine the temporal properties of Cyg X-1 by using 
data from high timing resolution modes ({\em Single-Bit} and {\em Event}). 
For each observation, we rebinned the {\em Event} data to $2^{-13}$ s 
time bins, which is the resolution of the {\em Single-Bit} data, and the 
combined all the data to cover the entire PCA passing band. We broke the 
light curve into 128 s segments and carried out a $2^{20}$-point 
{\em Fast Fourier Transformation} of each segment to obtain the 
corresponding power density spectrum (PDS) that is Leahy--normalized 
(Leahy et al. 1983). The power spectra of all segments were then weighted
and coadded to obtain the average PDS, as well as its variances. From 
the average PDS we subtracted the dead-time corrected power due to 
Poisson statistical 
fluctuation (see Zhang et al. 1995 for discussion on dead-time effects).
Finally, we divided the resulting PDS by the mean source rate to express
Fourier powers in terms of fractional rms amplitudes (van der Klis 1995).
The results are presented in Fig.~3 for the same selected observations
(as in Table~1).

Over the duration of the last flare (see Fig.~1), the PDS of the source
evolved significantly. The shape of the PDS went from being 
``flat--topped'' at the beginning, which is characteristic of the hard 
state, to being roughly $1/f$ at the peak, which is characteristic of the 
soft state; it returns to the flat-topped profile at the end, completing 
a full transition. In between, however, the PDS shows a mixture of the 
flat--topped component and the $1/f$ component, which is characteristic of 
the transitional periods between the hard and soft states (Cui et al. 
1997a; Cui et al. 1997b). There are clearly additional features. For 
the third and fourth observations, the PDS further steepens at roughly 
10 Hz, which was observed in the 1996 soft state (Cui et al. 1997a). For 
other observations, there is a broad but localized feature centered at a 
few Hz, again similar to what has been observed during the hard state 
(e.g., Belloni \& Hasinger 1990) and transitional periods in 1996 (Cui 
et al. 1997a). Fig.~3 also shows the overall variability of the source 
at different stages of the flare. The fractional rms amplitude shows a 
decreasing trend toward the peak, although the change is quite small
compared to that for transient BHCs during a state transition (van der
Klis 1995). We conclude that the flare is also similar to the 1996 state
transition in terms of the observed temporal properties of Cyg X-1.

\section{Summary and Discussion}

We observed Cyg X-1 during an unusual period when three consecutive major
X-ray flares occurred over a period of about five months. Following each
flare, the source did seem to completely return to the hard state before
the next one started, as indicated by the flux-hardness correlation. The
correlation also indicates that Cyg X-1 entered a sustained period during 
the last flare that is very similar to the 1996 soft state. This is 
supported by striking similarities between the observed spectral and 
temporal properties of the source during that time period and those 
during the 1996 soft state (although there are some differences). 
Therefore, we argue that the physical process that is responsible for 
triggering a flare or a state transition is likely to be the same. The 
difference between the two types of phenomena seems mostly quantitative.
For instance, the soft state lasted only for about a week during the 
flare, compared to more than two months in the 1996 episode
This difference might be due to the difference in the change of physical 
quantities, such as the fraction of accretion energy dissipated in the 
disk. 

Although only an empirical model was used for spectral studies, the 
results do clearly reveal the evolution of the source, thanks to the
improved coverages of the transitions. It was known previously that a 
broken power law would be required to model the spectrum of Cyg X-1 at
high energies (Ebisawa et al. 1996; Cui et al. 1997a), as opposed to a 
simple power law for most BHCs (Tanaka \& Lewin 1995). Not only did we 
confirm this in our investigation, we also quantified the evolution of 
this component throughout the transitions. Moreover, we confirmed the 
presence of a ``settling period'' after the low-to-high transition or
before the high-to-low transition, as suggested by Cui et al. (1997a), 
when the soft flux is at the soft-state level but the spectral and 
timing properties are still intermediate between the hard and soft 
states. Such a period manifests itself prominently in the PDS shape.
Finally, throughout the flare the observed 5--200 keV flux does not
vary significantly. This is again consistent with previous results 
from a study of the 1996 state transition (Zhang et al. 1997). 

We conclude that we can perhaps learn a great deal about the origin of 
rare state transitions of Cyg X-1 from studying more frequently occurring 
X-ray flares. We have tried to establish a connection between the two 
types of phenomena in this investigation. The next step will be to 
carry out, in a more systematic manner, detailed comparisons between 
a range of X-ray flares and state transitions (Feng, Ertmer, \& Cui, in 
preparation). 

\acknowledgments
This work was supported in part by NASA through grants NAG5-9098 and
NAG5-9998.

\clearpage

\begin{deluxetable}{lcccccccc}
\scriptsize
\tablecolumns{9}
\tablewidth{0pc}
\tablecaption{Key Spectral Parameters for Selected Observations}
\tablehead{
 & & \multicolumn{3}{c}{broken power-law\tablenotemark{1}} & \multicolumn{2}{c}{high-energy cutoff\tablenotemark{2}} \\
\cline{3-5} \cline{6-7} \\
\colhead{No.} & \colhead{Obs. ID\tablenotemark{3}} & \colhead{$\alpha_1$}& \colhead{$\alpha_2$} & \colhead{$E_b$} & \colhead{$E_c$} & \colhead{$E_f$} & \colhead{$\chi^2_{\nu}/dof$} & \colhead{Flux\tablenotemark{4}} \\
 & & & & (keV) & (keV) & (keV) & &  } 
\startdata
1 & 03-13-00 & $1.98^{+0.03}_{-0.08}$ & $1.60^{+0.01}_{-0.02}$ & $10.5^{+0.4}_{-0.2}$ & $22^{+1}_{-2}$ & $152^{+7}_{-8}$ & $1.05/232$ & $3.07$ \\
2 & 01-05-01 & $2.51^{+0.02}_{-0.04}$ & $1.81^{+0.01}_{-0.02}$ & $10.7^{+0.1}_{-0.1}$ & $24^{+1}_{-1}$ & $130^{+6}_{-7}$ & $1.17/229$ & $2.05$ \\
3 & 01-07-00 & $2.66^{+0.02}_{-0.03}$ & $2.00^{+0.01}_{-0.01}$ & $10.7^{+0.1}_{-0.1}$ & $21^{+1}_{-1}$ & $109^{+6}_{-4}$ & $1.32/278$ & $3.32$ \\
4 & 01-08-05 & $2.74^{+0.01}_{-0.03}$ & $1.96^{+0.01}_{-0.02}$ & $10.5^{+0.2}_{-0.1}$ & $22^{+1}_{-1}$ & $123^{+5}_{-6}$ & $1.21/228$ & $2.74$ \\
5 & 01-09-00 & $2.54^{+0.01}_{-0.05}$ & $1.90^{+0.01}_{-0.02}$ & $10.7^{+0.2}_{-0.1}$ & $21^{+3}_{-2}$ & $144^{+12}_{-14}$ & $1.16/228$ & $3.52$ \\
6 & 01-09-05 & $1.86^{+0.12}_{-0.06}$ & $1.55^{+0.02}_{-0.02}$ & $10.6^{+0.6}_{-0.4}$ & $19^{+2}_{-2}$ & $125^{+8}_{-8}$ & $1.00/236$ & $2.45$ \\
\tablenotetext{1}{$\alpha_1$ and $\alpha_2$ are soft and hard power-law photon indices, respectively, and $E_b$ is the break energy.}
\tablenotetext{2}{$E_c$ and $E_f$ are cutoff energy and e-folding energy, respectively.}
\tablenotetext{3}{The prefix ``50109-'' is omitted.}
\tablenotetext{4}{It is the observed 5-200 keV flux (in units of $10^{-8}\mbox{ }erg\mbox{ }cm^{-2}\mbox{ }s^{-1}$).} 
\enddata
\end{deluxetable}

\clearpage
\begin{figure}[t]
\epsfxsize=400pt \epsfbox{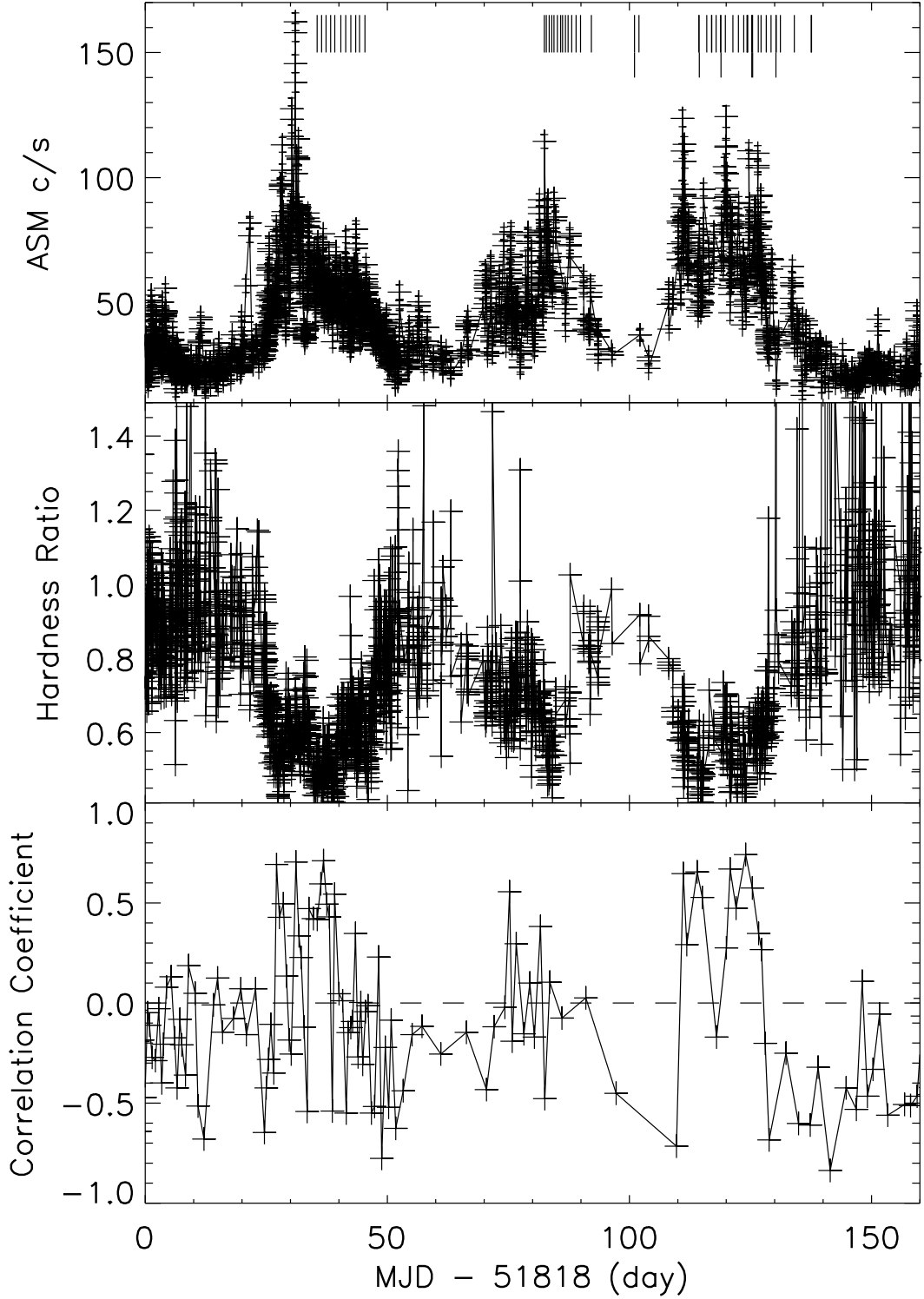}
\caption{(a) The ASM light curve (1.5-12 keV) of Cyg~X-1. It comprises 
measurements from individual ``dwells'' with 90-second exposure time.
The vertical lines at the top mark the times of the pointed observations,
with the longer ones indicating those that were selected for Table 1.  
(b) The time series of the ASM hardness ratio (3-12 keV/1.3-3 keV). (c) The 
correlation coefficient between the flux and the hardness ratio (see text). }
\end{figure}

\begin{figure}[t]
\psfig{figure=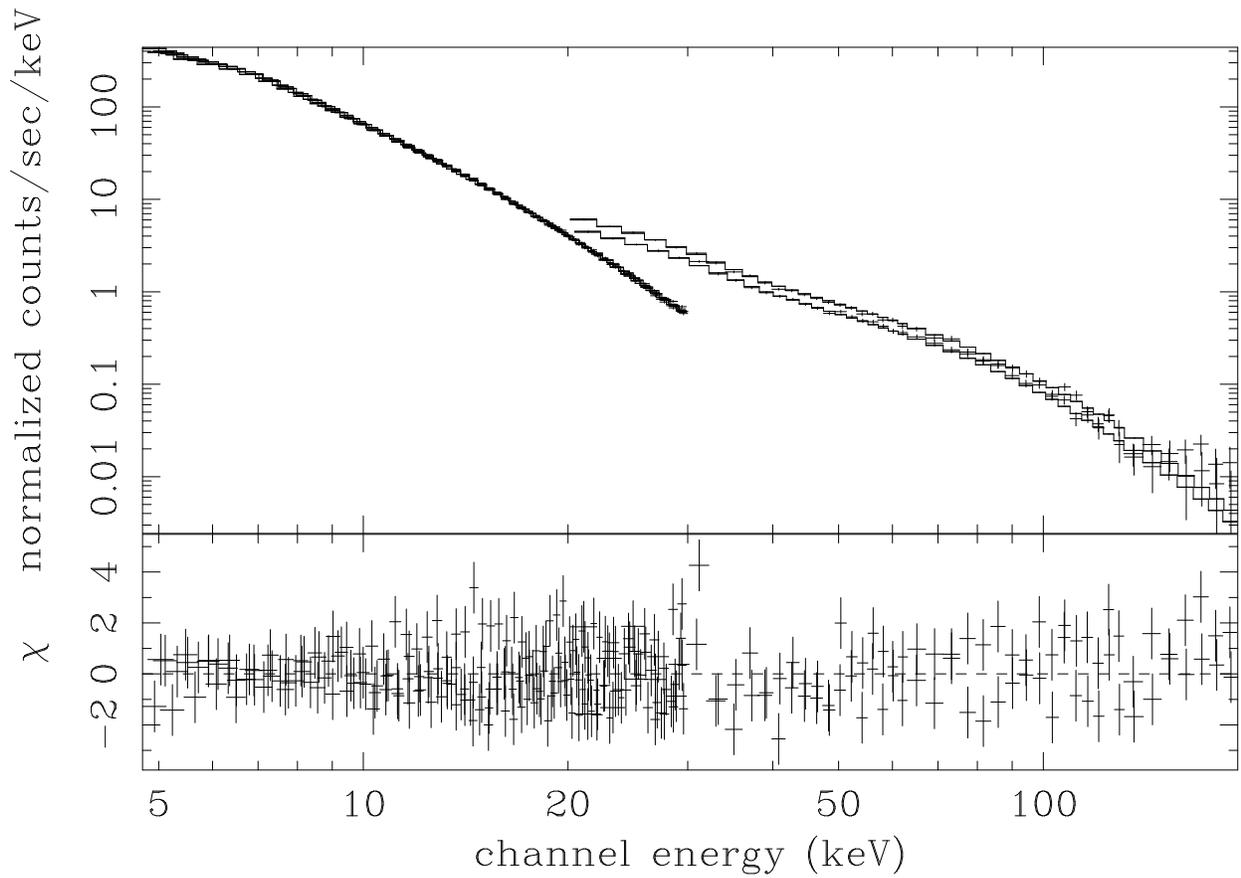,width=0.7\textwidth,angle=-90}
\caption{The combined PCA/HEXTE spectrum of Cyg~X-1 for the third observation 
in Table 1. The best-fit model is shown in solid histogram. The bottom panel 
shows the residuals of the fit.}
\end{figure}

\begin{figure}[t]
\epsfxsize=400pt \epsfbox{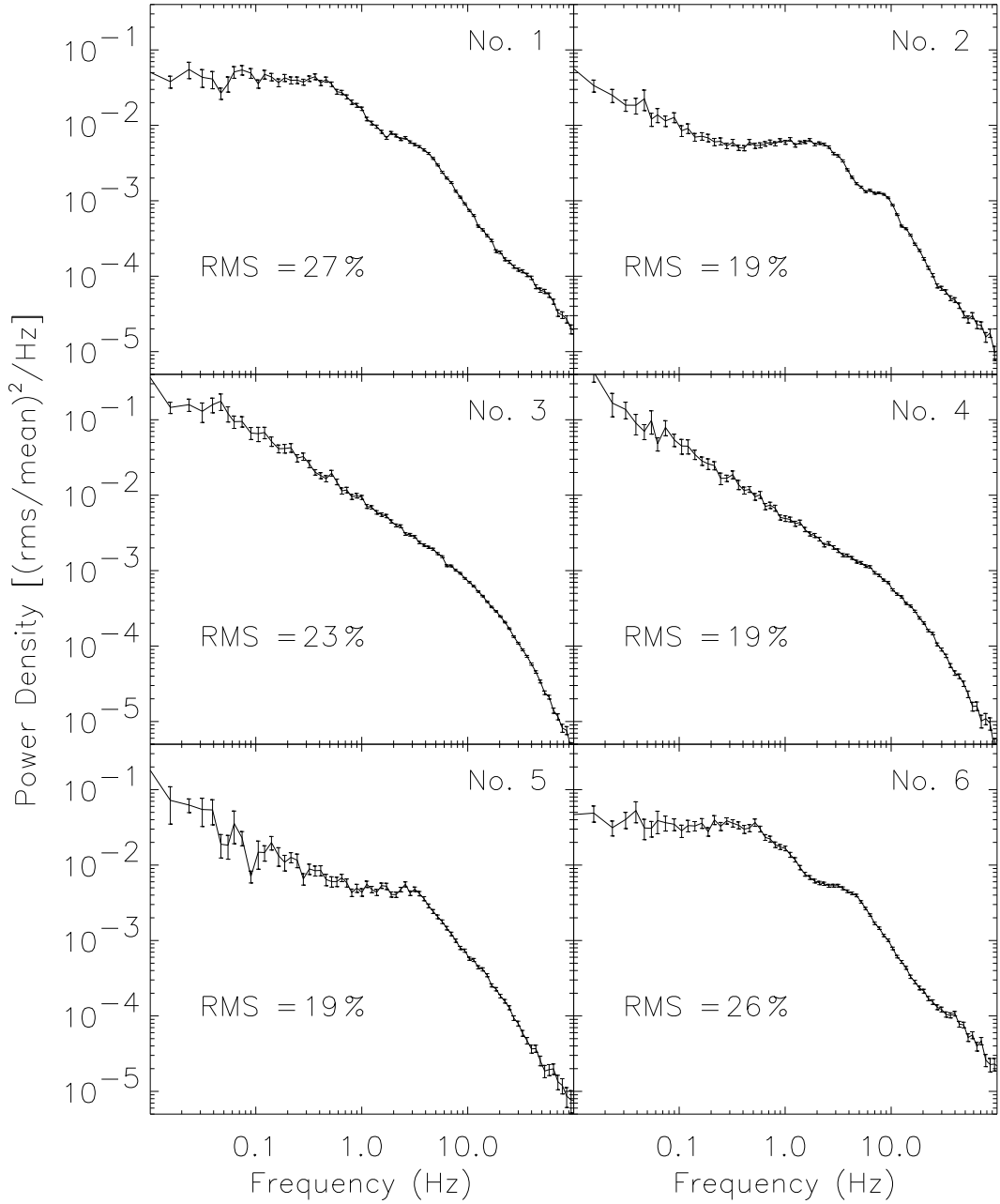}
\caption{The power density spectra of Cyg X-1 for a select sample of 
observations (see Table 1). The overall fractional rms amplitude 
(0.02 - 32 Hz) is also shown for each observation.}
\end{figure}

\end{document}